\documentclass[aps,showpacs,amsmath,superscriptaddress,twocolumn,prx]{revtex4-2}

\usepackage[nottoc,numbib]{tocbibind}

\usepackage[colorlinks=true, citecolor=black, linkcolor=black, urlcolor=black]{hyperref}

\usepackage[all]{hypcap}
\usepackage{amsmath}
\usepackage{enumerate,bm}
\usepackage{graphicx}
\usepackage{grffile} 
\usepackage{color}
\usepackage{esint}
\usepackage{textpos}
\usepackage[usenames,dvipsnames]{xcolor}
\usepackage{flushend}
\usepackage{tabularx}
\usepackage{amssymb}
\usepackage{float}
\usepackage{subfigure}
\usepackage[normalem]{ulem} 

\usepackage{lineno} 
\usepackage{setspace}
\usepackage{verbatim}
\usepackage{ragged2e}
\usepackage{url}

\floatstyle{ruled}
\newfloat{Box}{thp}{lop}

\begin{document}

\author{Alberto Castro}
\email{acastro@bifi.es}
\affiliation{Institute for Biocomputation and Physics of Complex Systems, University of Zaragoza, 50018 Zaragoza, Spain}
\affiliation{ ARAID Foundation, 50018 Zaragoza, Spain}

\author{Umberto~De~Giovannini}
\email{umberto.de-giovannini@mpsd.mpg.de}
\affiliation{Universit\`a degli Studi di Palermo, Dipartimento di Fisica e Chimica—Emilio Segr\`e, via Archirafi 36, I-90123 Palermo, Italy}
\affiliation{Max Planck Institute for the Structure and Dynamics of Matter and Center for Free Electron Laser Science, 22761 Hamburg, Germany}

\author{Shunsuke~A.~Sato}
\email{ssato@ccs.tsukuba.ac.jp}
\affiliation 
{Center for Computational Sciences, University of Tsukuba, Tsukuba 305-8577, Japan}
\affiliation{Max Planck Institute for the Structure and Dynamics of Matter and Center for Free Electron Laser Science, 22761 Hamburg, Germany}

\author{Hannes~H\"ubener*}
\email{hannes.huebener@mpsd.mpg.de}
\affiliation{Max Planck Institute for the Structure and Dynamics of Matter and Center for Free Electron Laser Science, 22761 Hamburg, Germany}

\author{Angel Rubio}
\email{angel.rubio@mpsd.mpg.de}
\affiliation{Max Planck Institute for the Structure and Dynamics of Matter and Center for Free Electron Laser Science, 22761 Hamburg, Germany}
\affiliation{Center for Computational Quantum Physics (CCQ), The Flatiron Institute, 162 Fifth avenue, New York NY 10010.}

\title{Floquet engineering the band structure of materials with optimal control theory}

\begin{abstract}
We demonstrate that the electronic structure of a material can be deformed
into Floquet pseudo-bands with arbitrarily tailored shapes. We achieve this goal 
with a novel combination of quantum optimal control theory and Floquet engineering.
The power and versatility of this framework is demonstrated here by utilizing the independent-electron 
tight-binding description of the $\pi$ electronic system of
graphene. We show several prototype examples focusing on the region around the K
(Dirac) point of the Brillouin zone: creation of a gap with opposing
flat valence and conduction bands, creation of a gap with opposing
concave symmetric valence and conduction bands -- which would
correspond to a material with an effective negative electron-hole mass --, or
closure of the gap when departing from a modified graphene model with
a non-zero field-free gap. We employ time periodic drives with several
frequency components and polarizations, in contrast to the usual monochromatic fields,
and use control theory to find the amplitudes of each component that
optimize the shape of the bands as desired. In addition, we use
quantum control methods to find realistic switch-on pulses that 
bring the material into the predefined stationary Floquet band structure, 
i.e. into a state in which the desired Floquet modes of the target bands
are fully occupied, so that they should remain stroboscopically stationary, with long lifetimes,
when the weak periodic drives are started. Finally, we note that although
we have focused on solid state materials, the technique that we
propose could be equally used for the Floquet engineering of ultracold
atoms in optical lattices, and to other non-equilibrium dynamical and correlated systems.
\end{abstract}

\maketitle

\section{Introduction}

The possibility of preparing materials in non-equilibrium steady states with tailored properties by exciting them with continuous lasers, and thereby manipulating -- even designing -- their electronic band structure has attracted considerable experimental and theoretical interest. Floquet theory maps the time-dependence of any periodically driven system into a structure of states in energy space that are observable as replica bands in time-resolved angular resolved photoelectron spectroscopy~\cite{Wang:2013fe,Mahmood:2016bu,Aeschlimann:2021hq}. A number of theoretical works \cite{Fregoso:2013di,Sentef:2015jp,deGiovannini:2016cb,Giovannini:2020hm,deGiovannini:2022ck} have proposed to interpret these dressed states or sidebands of such laser driven materials as new quasiparticle levels, and identify the non-equilibrium steady states with the eigenstates of the Floquet operator. It is thus possible to predict a variety of new properties of such driven material states by analyzing the Floquet eigenstates. This has led to the idea sometimes referred to as Floquet engineering: tailoring the material properties with a periodic drive, by manipulating the Floquet electronic band structure. In particular, the topology of electronic Floquet states has been discussed in many works. The paradigmatic example is graphene under circularly polarized irradiation, which has been theoretically predicted~\cite{oka_photovoltaic_2009, Kitagawa:2010bu,Inoue:2010iz} to attain the properties of a Chern insulator, a behavior which has been partially confirmed by ultrafast electronic pump-probe measurements~\cite{2018McIver,Sato2019a}. The manipulation of other Floquet topological phases~\cite{Wang:2014ji,Chan:2016dqa,Chan:2016ir,Ebihara:2016de,Yan:2016eea,Narayan:2016jl,Hubener:2017ht,Yan:2017bv,Ezawa:2017gv,Wang:2018us,Weber:2021fv,Titum2016}, and the engineering of topological bands ~\cite{lindner_floquet_2011,Ezawa:2013vm,Katan:2013hl,Narayan:2015vd,Claassen:2016ge,DAlessio:2015td,Dutreix:2016ck,Wang:2016ur,Zhang:2016di,Roy:2017vi,Liu:2018dk,Nguyen:2021uc} have also been proposed. 

Thus far, however, this field has fallen short on two important points: (i) In order to create a material with a given property, one should not only ensure that the Floquet bands have a particular shape, but also ensure that the electrons occupy the relevant bands. For example, the topological character, classified by the Chern number, is given by the integrated Berry curvature of the bands, which depends on the occupations of the states. (ii) The touted promise of band engineering has been limited to creating replica bands with hybridization gaps\cite{Wang:2013fe,Mahmood:2016bu}, instead of designing or largely modifying band structures in a stricter sense. 

In this work we address these two points by demonstrating the design of laser pulses capable of both creating Floquet band structures with predefined dispersion shapes, and fully controlling their occupations. These two properties, occupation and shape, are in fact controlled by two different pulses. The occupation is determined by how a first pulse connects the equilibrium ground state to the Floquet state, that is to say by how it is switched on. While some works take note of this fact, and simulate the switch-on phase~\cite{Kalthoff:2018et}, and others have analyzed the role of dissipation in the stabilization of Floquet phases~\cite{Dehghani:2014jm,Dehghani:2015gz,Seetharam:2015wj}, most works make strong a priori assumptions or approximations on the occupations of Floquet states. We note that a system in a single Floquet state in principle does not absorb energy from the periodic driving laser, which means that it can remain longer in the Floquet state before heating up in the long time limit. Therefore, driving protocols such as the ones discussed here may lead to an enhancement of the pre-thermal steady state~\cite{Weidinger:2017tb,Haldar:2018uz} lifetime, and thus to the experimental accessibility of Floquet state properties. 

The shape of the Floquet band structure is instead determined by the varying amplitudes and phases of the different frequencies of the continuous driving pulse that comes after the switch-on has been completed. It is important to note that the only condition for the Floquet theorem to apply is that the system Hamiltonian is periodic in time. There is no restriction on the number of frequencies that can be contained in the pulse, beyond the fact that they must be multiples of the fundamental frequency that determines its overall periodicity.

We will show that both the band shapes and the band occupations can be fully controlled by designing the driving field with optimal control theory. Specifically, we demonstrate, as a proof of principle, how one can design realistic multicolor driving fields that result in pre-defined shapes of Floquet bands in parts of the Brillouin zone (BZ), thereby fulfilling the promise of band engineering. One can achieve in this way almost arbitrary band shapes defined as control targets -- and for example control the effective masses, that can be given a negative, zero, or positive value. We further show that it is possible to design light fields such that a single Floquet band is selectively occupied, thus preparing the system in a single Floquet eigenstate. In this way, one may prepare a steady-state electronic structure that is entirely governed by the properties of that Floquet state.

Finally, we note that by shaping the band structure one is effectively shaping the nature of electron correlations (e.g., flat bands
are connected in many cases to strong electron correlation effects and other rich phenomena). 
It is an intriguing possibility to study the manipulation
of laser induced electronic correlations with optimal control, that we however leave for future work.

In Section~\ref{sec:model} we present the graphene model that we have used, with and without the presence
of time-dependent periodic perturbations, that can then be described with the help of Floquet formalism. Section~\ref{sec:method}
presents the combination of optimal control theory and the Floquet formalism that we propose in order
to perform Floquet engineering of the materials. Section~\ref{sec:results} shows results for some
band engineering possibilities: band flattening, band curvature inversion, gap closing. Finally, Section~\ref{sec:conclusions}
summarizes our conclusions.

\section{Model}
\label{sec:model}

In the following we will first briefly describe the tight-binding model of graphene that we have used, and its behavior under circularly polarized laser pumping, using Floquet theory. This will serve as the reference system for this proof of principle work. 

\begin{figure}
    \centering
    \includegraphics[width=\columnwidth]{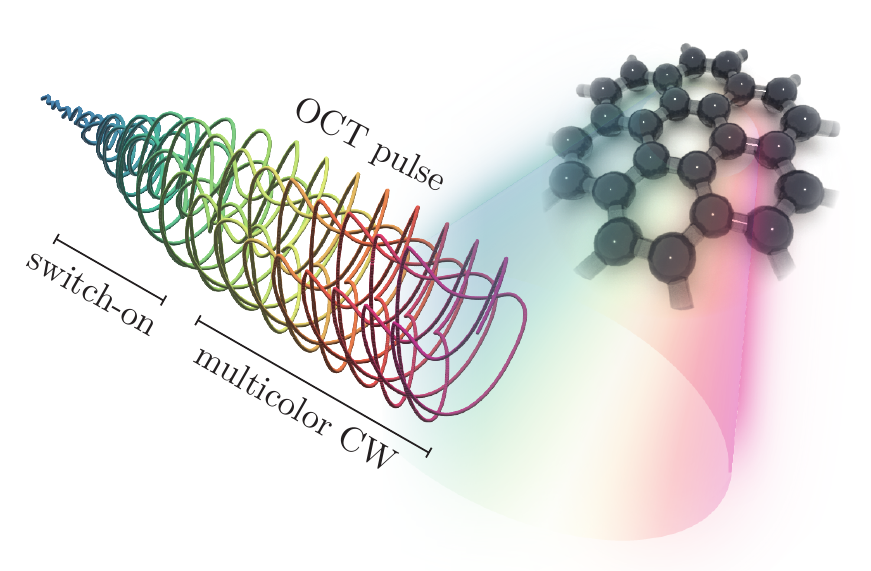}
    \caption{ Schematic diagram of the technique demonstrated in this article: A graphene
      sheet is first pumped by a switch-on pulse, that is designed by
      OCT techniques to prepare the system such that the Floquet bands
      induced by a periodic multicolor continuous wave pulse, that
      comes later, are populated. This time periodic driving is also
      designed by OCT methods, in order to produce Floquet bands with
      a given predefined shape.}
\end{figure}

We demonstrate the practical feasibility of true Floquet band engineering using graphene as an example, because it has been widely discussed in this context. We use 
this simple model of independent electrons that describes the low-energy states of graphene~\cite{haldane_model_1988,CastroNeto:2009wq}:
\begin{equation}\label{H0}
    H_{\mathbf{k}} = -\tau \left(\begin{array}{cc}
        0 & \Delta(\mathbf{k}) \\
        \Delta^*(\mathbf{k}) & 0
    \end{array}\right)  
\end{equation}
where
\begin{equation}
\Delta(\mathbf{k}) = e^{-i k_x a}\left[ 1+2 e^{i3k_x a/2}\cos\left(\frac{\sqrt{3}}{2}k_y a\right) \right]\,,
\end{equation}
with $a=2.68$ a.u. being the lattice constant, and $\tau=0.127$ a.u. This model parametrizes the $\pi$ bands of graphene, containing the linear dispersing Dirac regions around the $K$-points of the Brillouin zone. Upon irradiation with an electric field (to be treated in what follows in the long wavelength or dipole approximation), this Hamiltonian must be modified to include an applied time-dependent vector potential $\mathbf{A}(t)$; this can be achieved with the substitution $\mathbf{k} \to \mathbf{k}+\mathbf{A}(t)$. 

For a vector potential with periodicity $T=2\pi/\Omega$, Floquet theory can be applied.
We start by defining the Floquet bands $\epsilon_{\mathbf{k}\alpha}$ (also known as pseudoenergies, or pseudobands),
in terms of the eigenvalues
of the evolution operator at the periodic time $T$:
\begin{equation}
\label{eq:floquetmodes}
U_\mathbf{k}(T)\vert u^\alpha_\mathbf{k}\rangle = e^{-i\epsilon_{\mathbf{k}\alpha}T}\vert u^\alpha_\mathbf{k}\rangle\,.
\end{equation}
Because, in our case, the basis space of the model is two-dimensional, there are only two eigenvalues per $k$-point ($\alpha = 0, 1$).
However, note that the pseudobands are only defined modulo $\Omega$, such that all $\epsilon_{\mathbf{k}\alpha}^{(n)} = \epsilon_{\mathbf{k}\alpha} + n\Omega$
(for all $n \in \mathbb{Z}$) are valid values. These replicas of the pseudobands are often called sidebands.
One may also define a so-called ``effective'' Hamiltonian, that verifies
\begin{equation}
U_\mathbf{k}(T) = e^{-i H^{\rm eff}_\mathbf{k} T}\,,
\end{equation}
It must have the same eigenstates,
$H^{\rm eff}_\mathbf{k}\vert u^\alpha_\mathbf{k}\rangle = \epsilon_{\mathbf{k}\alpha}\vert u^\alpha_\mathbf{k}\rangle$,
also called the Floquet modes. These
can be used to expand any solution to Schr{\"{o}}dinger's equation as:
\begin{equation}
\label{eq:floquettheorem}
    \vert\psi_{\mathbf{k}}(t)\rangle = \sum_\alpha f_{\mathbf{k}\alpha} e^{{i\epsilon_{\mathbf{k}\alpha}t}}
\vert u^\alpha_{\mathbf{k}}(t)\rangle\,.
\end{equation}
In this equation, the time-dependent Floquet modes $\vert u^\alpha_{\mathbf{k}}(t)\rangle$ are defined
from the following propagation of the static modes $\vert u^\alpha_{\mathbf{k}}\rangle$ 
(that were defined through Eq.~(\ref{eq:floquetmodes})):
\begin{equation}
\vert u^\alpha_{\mathbf{k}}(t)\rangle = e^{-i\epsilon_{\mathbf{k}\alpha}t}U_\mathbf{k}(t) \vert u^\alpha_{\mathbf{k}}\rangle\,.
\end{equation}
These modes are time-periodic.
The complex coefficients $f_{\mathbf{k}\alpha}$ describe how much each Floquet state is 
contributing to the dynamics, and hence we define them as the occupation of the Floquet states.
Floquet theory permits to simplify the treatment of periodically driven systems in the following
way: once the evolution of the Floquet modes $\vert u^\alpha_{\mathbf{k}}(t)\rangle$ in a single period is known (the
so-called ``micromotion''), the long term evolution of any state can be easily obtained via Eq.~(\ref{eq:floquettheorem}).

For practical purposes, it is useful to decompose these time-periodic Floquet modes into their Fourier components:
\begin{align}
\label{eq:fourier1}
\vert u^\alpha_{\mathbf{k}}(t)\rangle &= \sum_{m=-\infty}^{\infty}
 e^{i m \Omega t} \vert u^\alpha_{\mathbf{k}m}\rangle\,,
\\
\label{eq:fourier2}
\vert u^\alpha_{\mathbf{k}m}\rangle\ &= \frac{1}{T} \int_0^T\!\!{\rm d}t\; e^{-im\Omega t} \vert u_{\mathbf{k}\alpha}(t)\rangle\,,
\end{align}
that can then be found using the following eigenvalue equation (that must be truncated at some finite harmonic component value):
\begin{equation}
\sum_{n}\mathcal{H}^{mn}_\mathbf{k} |u^\alpha_{\mathbf{k}n}\rangle = \epsilon_{\alpha\mathbf{k}} |u^\alpha_{\mathbf{k}m}\rangle\,,
\end{equation}
\begin{equation}
\mathcal{H}_\mathbf{k}^{mn}  = \frac{\Omega}{2 \pi}\int_{2\pi/\Omega} dt e^{i(m-n)\Omega t} H_\mathbf{k}(t) 
+ \delta_{mn}m\Omega\,,
\end{equation}
where $m$ and $n$ label the harmonic components of each Floquet eigenstate, and $\alpha$ labels the pseudoband.
These Fourier modes, that can be identified with the sidebands mentioned above, correspond to the observed sideband Floquet slates in ARPES experiments.

\begin{figure*}\label{fig:effectiveMass}
    \centering
    \includegraphics[width=\textwidth]{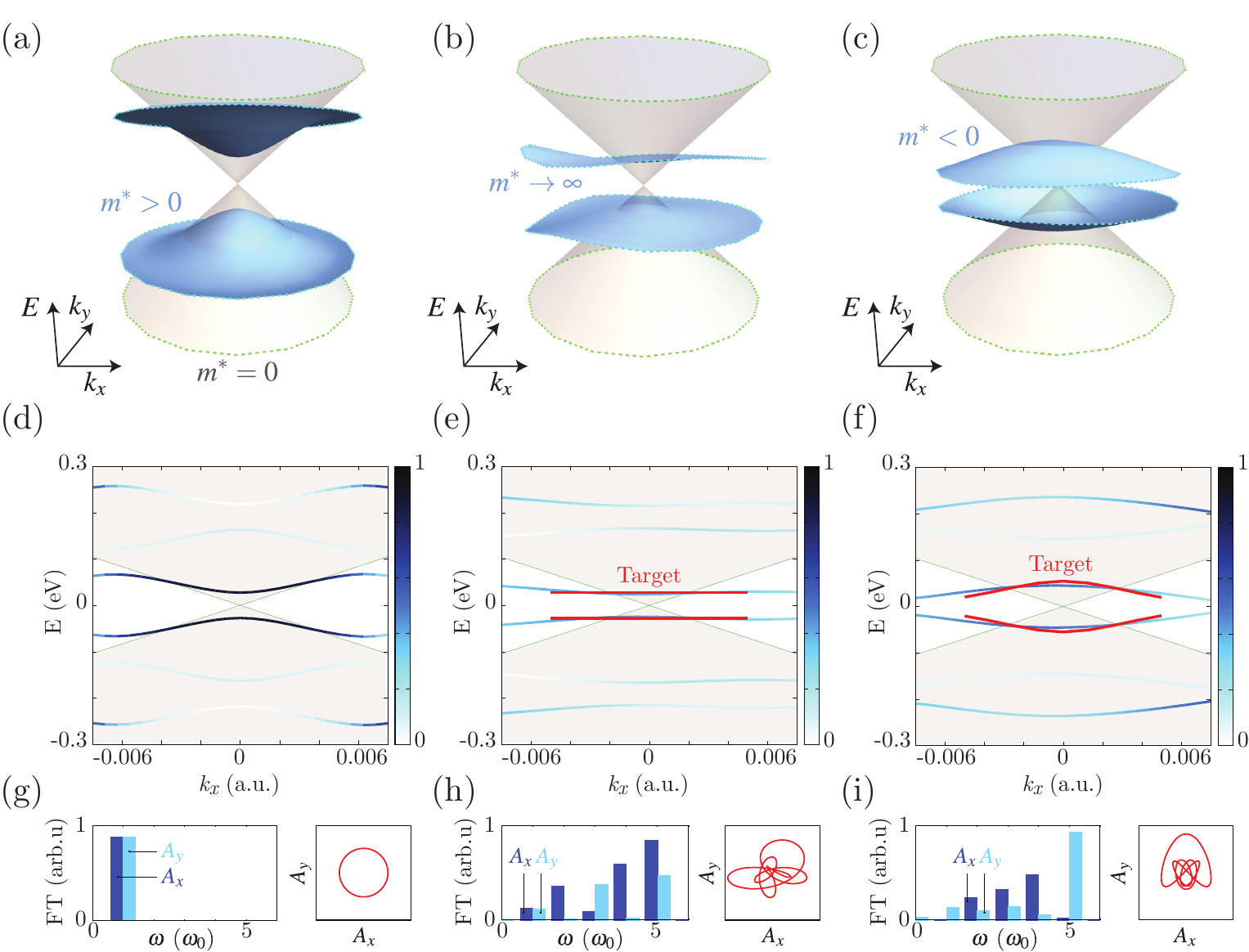}
    \caption{ {\it Top panels (a), (b), and (c):} Field-free bands
      (grey) and Floquet pseudo-bands (blue) of graphene, in the
      vicinity of the K point.  The Floquet pseudo-bands are computed
      (a) using a reference circularly polarized field (see text); (b)
      using fields optimized to create a gap with opposing parallel
      bands, and (c) using fields optimized to create a gap with
      opposing concave bands. {\it Middle panels (d), (e), and (f):}
      Cuts through the $xz$ plane of the previous 3D plots, displaying
      also the target bands in red, and the series of Floquet
      replicas. These are colour-coded: a darker blue color
      corresponds to a larger Fourier component $\langle
      u^\alpha_{\mathbf{k}m} \vert u^\alpha_{\mathbf{k}m}\rangle$ [see
        Eqs.~(\ref{eq:fourier1}) and (\ref{eq:fourier2})]. {\it Bottom
        panels (g), (h), and (i):} Fourier decomposition (left) and
      Lissajous plots (right) of the fields corresponding to
      the pseudo-bands shown above.}
\end{figure*}

The recently most widely discussed~\cite{2018McIver,Jotzu:2014kz,Burkov:2011uv,Oka:2018df,Rudner:2020tq,Bao:2022wl} case of Floquet engineering is the possibility to endow a material with topological properties by applying circularly polarized light, in particular in the case of graphene. Fig.~\ref{fig:effectiveMass}(a) shows the Floquet bands at the Dirac point of graphene under circularly polarized illumination with a 6.5 $\mu$m wavelength ($\approx$ 190 meV), and an intensity of 20~MV/m. The linear crossing of the Dirac bands is deformed in the Floquet steady state to admit a band gap (the field-free linearly crossing bands are also shown for reference). The series of replica bands is shown for a cut across the Dirac point in Fig.~\ref{fig:effectiveMass}(d). These sidebands are color-coded: the blue intensity is proportional to $|u^\alpha_{\mathbf{k}m}\vert$, which represents the contribution of the corresponding component in the Fourier expansion to the Floquet state. It can be noted that, with increasing harmonic number $|m|$, the contributions become negligibly small.

The characteristic gap opening shown in Fig.~\ref{fig:effectiveMass}(a) is, for large frequencies, inversely proportional to the frequency of the applied light, and directly proportional to the intensity~\cite{oka_photovoltaic_2009} (although this dependence is only valid in the high frequency limit). The analysis of the Berry curvature of the Floquet modes (see Fig.~\ref{fig:berry}) shows that they are non-zero and integrate to non-trivial Chern numbers. Defined as the BZ integral of the Berry-curvature of all occupied bands, the Chern number is a topological invariant, and is connected to the observation of the anomalous quantum Hall effect~\cite{haldane_model_1988,Kane:2005uo,CuiZu:2013dd}.

While each of the two Floquet bands of graphene under circularly polarized illumination separately integrates to a Chern number, they have opposite signs: unless one is completely empty and the other one is completely occupied, the overall system is not in a topological phase. Instead, its steady state would be characterized by a mixture of opposite Chern numbers. Hence, the creation of novel Floquet phases requires control over the population of Floquet states. The band engineering with the circularly polarized light in graphene shown in Fig.~1(b) is limited to an opening of a gap at the Dirac point, which may be controlled by the intensity of the light, but does not address the problem of preparing the system in a state with the appropriate occupations. We will now discuss how optimal control theory (OCT) can be utilized to design pulses that both control the occupations, as well as the shape of the Floquet bands.

\section{Method: Quantum Optimal Control Theory (QOCT)}
\label{sec:method}

Optimal control theory addresses the following mathematical problem~\cite{Kirk1998,Shapiro2003,Brif2010,Castro2018}: given a dynamical
system (e.g. Schr{\"{o}}dinger's equation) that can be controlled with a set of {\it control} parameters (e.g. the frequencies and
amplitudes of an external electromagnetic field), find those parameters that optimize the behavior of the system with respect to the
achievement of some predefined target (e.g. the population of some given state at the end of the process).  We are
concerned here with two different target definitions: (I) given a material subject to periodic fields, find the temporal shape of those
fields such that the generated Floquet pseudo-bands most closely resemble some predetermined dispersion,
and (II) given the same material in
its ground state, find the shape of the switch-on fields that drive the electrons in a valence band to occupy a given Floquet
pseudo-band (in particular, the pseudo-band induced by the fields that have been optimized previously).

Let us start with a brief description of the method that we have used for target (I). 
The control parameters $u_1,\dots,u_P$ will be the Fourier coefficients that determine the shape of the vector potential ${\bf A}(t)$:
\begin{align}
\label{eq:afourier1}
  A_x(u, t) &= \sum_{n=1}^M u_{2n}\cos(\Omega_n t) + u_{2n-1} \sin(\Omega_n t)\,,
\\
\label{eq:afourier2}
  A_y(u, t) &= \sum_{n=1}^M u_{2M + 2n}\cos(\Omega_n t) + u_{2M + 2n-1} \sin(\Omega_n t)\,.
\end{align}
The frequencies $\Omega_n = \frac{2\pi}{T}n$ are multiples of the fundamental frequency $\Omega$ used
for the previously described example of graphene irradiated with circularly polarized light, and $T$ is
the corresponding period. Note that by using this parametrization we automatically include a {\it cutoff}
frequency $M\Omega$, which also sets the total number of control parameters ($4M$). 
The existence of a cutoff frequency is a natural experimental constraint, too. We assume the possibility
of independently shaping the fields in the $x$ and $y$ direction.

These fields determine the temporal shape of the Hamiltonian at each ${\bf k}$-point: at the low
amplitudes that we assume in this work, we can expand $H_{{\bf k} + {\bf A}(u, t)}$
in terms of the field ${\bf A}(u, t)$, and retain the linear part, i.e.:
\begin{equation}
H_{\bf k}(u, t) = H_{\bf k} + \sum_{i=x,y}A_i(u, t)H^i_\mathbf{k}\,.
\end{equation}
These Hamiltonians, in turn, determine the evolution operators:
\begin{align}
i\frac{\partial U_\mathbf{k}[u]}{\partial t}(t) &= H_\mathbf{k}(u, t)U_\mathbf{k}[u](t)\,,
\\
U_\mathbf{k}[u](0) &= I\,.
\end{align}
The Floquet pseudo-bands $\epsilon_{\mathbf{k}\alpha}(u)$ and the corresponding Floquet modes
$\vert u^\alpha_{\mathbf{k}}(u)\rangle$ can then be defined using the eigenvalues and eigenfunctions of the evolution operator
at the periodic time $T$:
\begin{equation}
U_{\bf k}[u](T) = \sum_\alpha e^{-i\epsilon_{\mathbf{k}\alpha}(u) T} 
\vert u^\alpha_{\mathbf{k}}(u)\rangle\langle u^\alpha_{\mathbf{k}}(u)\vert\,.
\end{equation}

Suppose now that we wish for the exteranl fields to induce a given set of 
pseudo-bands $\tilde{\epsilon}_{{\mathbf k}\alpha}$ and
Floquet modes $\vert \tilde{u}^\alpha_\mathbf{k}\rangle$. This would be equivalent to asking the perturbation
fields to induce an evolution operator given by:
\begin{equation}
\tilde{U}_{\bf k} = \sum_\alpha e^{-i\tilde{\epsilon}_{\mathbf{k}\alpha} T} 
\vert \tilde{u}^\alpha_{\mathbf{k}}\rangle\langle \tilde{u}^\alpha_{\mathbf{k}}\vert\,.
\end{equation}
One can use quantum optimal control theory for the generation
of target evolution operators. This concept was first developed for the problem of designing quantum gates
within quantum information theory studies~\cite{Palao2002}. However, we may use it for the problem at hand -- with the added
complication that the target must be simultaneously formulated for a set of ${\bf k}$-points.

Thus, the mathematical formulation of the optimization problem can be established by defining the following
target functional:
\begin{equation}
\label{eq:functionalF}
F(U_\mathbf{k}[u](T)) = \frac{1}{N_{\rm kps}}
\sum_{\mathbf{k} \in \mathcal{K}}\vert 
\tilde{U}_\mathbf{k} \cdot U_\mathbf{k}[u](T)
\vert^2\,.
\end{equation}
where we use the Fr{\"{o}}benius dot product for operators:
\begin{equation}
A \cdot B = \frac{1}{d} {\rm Tr} A^\dagger B\,. 
\end{equation}
$d$ is the dimension of the operators, which in our case is two.
$N_{\rm kps}$ is the number of $\mathbf{k}$-points in the set
$\mathcal{K}$: this is a finite set of points in the Brillouin zone,
that defines the region of interest: the goal is to engineer the band
structure in this region. Here, we will work with a disk-shaped region
defined around the $K$ point of graphene.

The functional in Eq.~(\ref{eq:functionalF}) takes a maximum value equal
to one when the generated evolution operators $U_\mathbf{k}[u](T)$ are
equal to the target ones $\tilde{U}_\mathbf{k}$ (or are equivalent,
i.e. related by a global multiplicative phase factor). The goal is
therefore finding the parameters $u$ that lead to those evolutions:
the problem finally boils down to the maximization of the
function:
\begin{equation}
\label{eq:g}
G(u) = F(U_\mathbf{k}[u](T))\,.
\end{equation}
Note that, in this work, we are only interested in the shape of the pseudo-bands $\epsilon_{\mathbf{k}\alpha}$, regardless
of the modes $\vert u_\mathbf{k}^\alpha\rangle$. However, in the formulation
described above, we need to specify the target modes $\vert \tilde{u}_\mathbf{k}^\alpha\rangle$. In this work, we
have simply set those modes to be equal to the field-free states associated to the corresponding band~\footnote{
Setting the target Floquet modes in the target functional adds an unnecessary extra constraint to the optimization process, if only
the energy bands are the target. This constraint could be lifted with a different formulation, written 
only in terms of the proximity of the generated pseudo-bands to the target ones. We will develop this idea in a
follow-up publication.
}.

In order to find the maximum of a multivariate function $G$, there is a plethora of available algorithms; the most efficient
ones require the use of the gradient $\nabla G(u)$. We have used the
the Sequential Least-Squares Quadratic Programming (SLSQP) algorithm~\cite{Kraft1994}
as implemented in the NLOPT library~\cite{nlopt}. Optimal control theory permits
to derive an expression for the gradient:
(essentially, as an application of Pontryagin's maximum principle~\cite{Pontryagin1962}). 
\begin{align}
\nonumber
\frac{\partial G}{\partial u_m}(u) &= \frac{1}{N_{\rm kps}} \sum_{i=x,y} \sum_{\mathbf{k}\in\mathcal{K}} 
\\\label{eq:gradient}
& 2{\rm Im}
\int_0^{t_f}\!\!\!{\rm d}t\; \frac{\partial A_i}{\partial u_m}(u, t)
B_\mathbf{k}[u](t)\cdot (H^i_{\mathbf{k}}(t)U_\mathbf{k}[u](t))\,,
\end{align}
where the {\it costate} $B_\mathbf{k}[u](t)$ is defined by the following equations:
\begin{eqnarray}
\label{eq:costate1}
i \frac{\partial}{\partial t}B_\mathbf{k}[u](t) &= H_\mathbf{k}(u, t) B_\mathbf{k}[u](t)\,,
\\
\label{eq:costate2}
B_\mathbf{k}[u](T) &= (\tilde{U}_\mathbf{k} \cdot U_\mathbf{k}[u](T) ) \tilde{U}_\mathbf{k}\,.
\end{eqnarray}
This equation of motion is analogous to the one that defines the evolution operator $U_\mathbf{k}[u](t)$,
except for the boundary condition, that is given at the final time of the propagation $T$: it is 
a {\it final} condition, instead of an initial condition.

Eqs.~(\ref{eq:gradient}), (\ref{eq:costate1}) and (\ref{eq:costate2}) enable us to compute the gradient
of the function $G$ defined in Eq.~(\ref{eq:g}), which allows its maximization, and the solution
of the first OCT problem mentioned above: (I) finding the periodic drivings that permit the 
Floquet pseudo-bands engineering. The second OCT problem -- (II) finding the switch-on
laser pulses that permit to populate those pseudo-bands starting from the material at equilibrium -- 
is a more conventional one. In this case, the problem and the target functional are not defined in terms of the evolution
operator of the system, but rather in terms of the material states.
The goal is to drive the electrons in the valence band towards the Floquet modes; the
states are driven by Sch{\"{o}}dinger's equation in the presence of a switch-on time-dependent
Hamiltonian:
\begin{align}
i\frac{\partial \psi_{\mathbf{k}\alpha}[v]}{\partial t}(t) &= H_\mathbf{k}(v, t)\psi_{\mathbf{k}\alpha}[v](t)\,,
\\
\psi_{\mathbf{k}\alpha}[v](0) &= \psi_{\mathbf{k}\alpha}\,,
\end{align}
where $\alpha$ is the valence band index ($\alpha = 0$, in our case).
The switch-on pulses are given by two vector field functions $A^{\rm so}_x(v, t)$ and $A^{\rm so}_y(v, t)$, that
operate a time $T_{\rm so}$, after which they are substituted by the optimal periodic driving that have been
found using the method described above. Once again, the shapes of the pulses are determined by a set of parameters
$v$, but in this case the parametrization is different. We have enforced the following shape:
\begin{equation}
\label{eq:switchonpulses}
A^{\rm so}_i(v, t) = S(t)f_i(v, t)A_i(u^{\rm opt}, t)\,,\quad i=x,y.
\end{equation}
Function $S(t) = 3(t/T_{\rm so})^2 - 2(t/T_{\rm s})^3$ smoothly varies from zero at $t=0$ to one
at $t=T_{\rm s}$. The fields $A_i(u^{\rm opt}, t)$ are the optimal periodic pulses that induce
the target Floquet pseudo-band structure. The functions $f_i(v, t)$ are, once again, Fourier series, similar
to the ones in Eqs.~(\ref{eq:afourier1}) and (\ref{eq:afourier2}), and the control parameters $v$ are
the corresponding Fourier coefficients. In this case, they are constrained to ensure that $f(u, t=T_{\rm so})=1$.
With this definition, the periodic pulses $A_i(u^{\rm opt}, t)$ are initiated, but are first multiplied by
a smooth envelope function $S(t)f_i(u, t)$ that slowly switches from zero to one at $t=T_{\rm so}$. After that switch-on
pulse time $T_{\rm so}$, the optimal periodic drivings stay indefinitely. The shape of the optimized
envelope should be such that, by the end of the switch-on phase, the ground state orbitals of
the material valence band are transformed into the Floquet modes and, in this way, they
are afterwards stationary (in the Floquet sense, that is, {\it stroboscopically} stationary).

It remains to specify the target functional for this case: if the goal is to maximize
the population, at time $T_{\rm so}$, of the Floquet modes $\vert u_{\mathbf{k}\alpha}\rangle$,
the natural choice is:
\begin{equation}
\label{eq:target2}
F(\psi_{\mathbf{k}\alpha}[v](T_{\rm so})) = 
\sum_{\mathbf{k}\in\mathcal{K}} \vert \langle \psi_{\mathbf{k}\alpha}[v](T_{\rm so})\vert u_{\mathbf{k}\alpha}\rangle\vert^2
\end{equation}
Function $G$ is now defined in terms of this new functional:
\begin{equation}
\label{eq:g2}
G(v) = F(\psi_{\mathbf{k}\alpha}[v](T_{\rm so}))\,,
\end{equation}
whose gradient is given by
\begin{align}
\nonumber
\frac{\partial G}{\partial v_m}(v) &= \sum_{i=x, y}\sum_{{\mathbf{k}} \in\mathcal{K}}
\\\label{eq:gradient2}
& 2{\rm Im}
\int_0^{t_f}\!\!\!{\rm d}t\; \frac{\partial A^{\rm so}_i}{\partial v_m}(v, t)
\langle \chi_{\mathbf{k}\alpha}[v](t)\vert H^i_{\mathbf{k}}(t)\vert \psi_{\mathbf{k}\alpha}[v](t)\rangle\,.
\end{align}
The costates $\chi_{\mathbf{k}\alpha}[v](t)$ are now defined by:
\begin{align}
\label{eq:costate3}
i \frac{\partial}{\partial t}\chi_{\mathbf{k}\alpha}[v](t) &= H_\mathbf{k}(v, t) \chi_{\mathbf{k}\alpha}[v](t)\,,
\\
\label{eq:costate4}
\chi_{\mathbf{k}\alpha}[v](T) &= \langle u_{\mathbf{k}\alpha}\vert\chi_{\mathbf{k}\alpha}[v](T)\rangle u_{\mathbf{k}\alpha}\,.
\end{align}
Eqs.~(\ref{eq:gradient2}), (\ref{eq:costate3}) and (\ref{eq:costate4}) permit to compute the gradient
of function $G$ defined in Eq.~(\ref{eq:g2}), which can then be fed into any function maximization algorithm
to solve the second optimization problem posed above.

A final important word about the computational methodology: the parameter search space is not unbound, as it would
allow for solution fields with arbitrary intensities, which would be experimentally impractical. We imposed
bound constraints, $\vert u_m \vert \le \kappa$, for some predefined bound $\kappa$.

\section{Results}
\label{sec:results}

\subsection{Band engineering}

We now describe the obtained results, starting with an
optimization of type (I): the first goal was to engineer a periodic
driving field that induces Floquet bands that are optimized to
resemble two parallel disks in a region around the Dirac point,
thus creating a gap in between two flat bands.
As the fundamental frequency $\Omega$, we used the same one (corresponding to
6.5 $\mu$m) used in the previously discussed reference example with circularly
polarized light. We set the cutoff frequency to five times $\Omega$,
i.e. $M$ = 5 in the parametrization given in Eqs.~(\ref{eq:afourier1})
and (\ref{eq:afourier2}). Furthermore, we establish a bound constraint
for the amplitudes of each individual frequency component ($\vert u_m
\vert \le \kappa$), where we set $\kappa$ to match the amplitude of
the reference circularly polarized example (20 MV/m).

The radius of the target disks was chosen to be 0.005 a.u., which
amounts to $\approx$ 0.3\% the length of a reciprocal lattice
vector. The manipulation of the Floquet bands shape, in these
examples, is therefore limited to a reciprocal space region around the
K point of that size. This size was chosen considering that we
are interested in some energy window around the Fermi level: the
energy separation of the field-free valence and conduction bands for
graphene around the K point (where the Fermi level is placed), at our
chosen radius of 0.005 a.u. is already of the order of the order of
0.1 eV [see Fig.~\ref{fig:effectiveMass}(c)], and thus, of the order
of 1000 K. This is the frequency range that must be used to search for
the optimal periodic drivings capable of manipulating the Floquet
bands. In the presence of frequencies below those temperatures, the
relevant region in reciprocal space is therefore the one that we have
used. We note that we have also successfully performed similar
optimizations using larger targer regions in reciprocal space -- at
the cost of using a correspondingly larger frequency range for the
periodic drivings.

Fig.~\ref{fig:effectiveMass}(b) shows the resulting bands together with the original Dirac
cones. Note that the target region, where the bands are optimized,
comprises only a part of the shown area, hence on the edges the disks
are slightly bent. Fig.~\ref{fig:effectiveMass}(e) shows a cut through \ref{fig:effectiveMass}(b) along the $k_x$
direction, and the target disks are indicated in red. We see here more
clearly that the OCT process results in bands that are almost
perfectly matching the target in the region where it is defined, and
the parallel character of the bands extends somewhat beyond that
region.

Parallel bands correspond to states that are well localized in real
space, and have been widely discussed in connection with twisted
bilayer systems, where the electron localization is caused by the interaction of the electronic structure at certain twist angles. 
Here instead, the driving pulse creates a superposition of valence and conduction states that result in a very similar band structure, but without the need for structural twisting and within the original Brillouin zone. 
The degree
of spatial localization is dictated by how much of the BZ is spanned
by the Floquet flat band. In our example in Fig.~\ref{fig:effectiveMass}(e) it is around one-percent, which is similar to
the localization length in moir\'e materials. Note that
this is different from the previously discussed dynamical
localization~\cite{Dunlap:1986co}, that results from a monochromatic
high-frequency drive, and which in the case of graphene only results
in the bands shown in Fig.~\ref{fig:effectiveMass}~(b). Instead, the
optimized pulse here has a complex frequency structure, as shown in
Fig.~\ref{fig:effectiveMass}~(h): it displays the absolute value of
its Fourier components (essentially, the control parameters $u$, see
Eqs.~(\ref{eq:afourier1}) and (\ref{eq:afourier2})). The fundamental
frequency for the flat Floquet band optimization is the same as for
the simple monochromatic Floquet case in
Fig.~\ref{fig:effectiveMass}~(a), and the gap opens by the same
amount, but the optimized pulse has additional
components. Furthermore, it has a non-trivial polarization dynamics,
as shown in the right panel of Fig.~\ref{fig:effectiveMass}~(h), where
the Lissajous plot of the electric field is shown (compare to the
simple circle in Fig.~\ref{fig:effectiveMass}~(g)).

Having demonstrated that it is possible to create flat bands in a defined region of the BZ, we take the band engineering one step further to demonstrate that, in principle, arbitrary band shapes can be designed. Fig.~1(d) shows a Floquet band structure where the effective mass of the bands has been inverted relative to the monochromatic case: instead of a convex band with a positive effective mass, it is possible to create a concave band where the effective mass is negative. Fig.~1(g) is a cut along 1(d), and it shows in red the target bands, along with the Floquet bands and sidebands, in varying tones of blue, once again graded according to their Fourier component magnitude. While for the present case there is no direct application of such band inversion (at least to our knowledge), in other materials it will result in drastically altered optically properties.

\begin{figure}
\centerline{\includegraphics[width = 0.67\columnwidth]{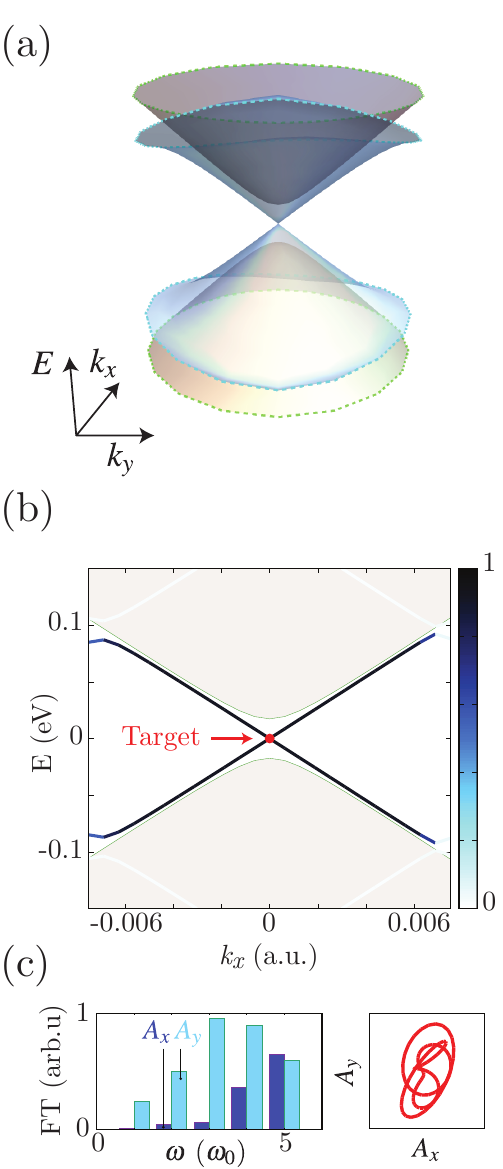}}
\caption{\label{fig:bandclosing1}
Gap closing.
(a) Field-free bands (grey) of the modified graphene model in the vicinity of the $K$ point (blue) showing
how the gap can be closed.
(b) Cut along the $y$ axis of panel (a). Along with the Floquet bands in the first BZ, the sidebands are also shown. The colour intensity of each point is proportional to the value of the corresponding Fourier component.
(c) Fourier components and Lissajou figure of the optimized pulse.
}
\end{figure}

For completeness, we have also attempted an optimization with the target of closing a gap, rather than opening it as in the previous cases. Since graphene does not have a gap, we have modified the graphene model that we have used, adding to the static Hamiltonian the term
\begin{equation}
    H^{\rm gap}_{\mathbf{k}} = \left(\begin{array}{cc}
        \delta/2 & 0 \\
        0 & -\delta/2
    \end{array}\right)\,,
\end{equation}
that creates a gap in the $K$ point. This modification corresponds to the
trivial part of the Haldane model~\cite{haldane_model_1988}.
We have not used the off-diagonal terms of the
Haldane model, that determine its topological properties, since those
would not change the discussion below about closing
the band gap.  For the value of the gap $\delta$ we have chosen 0.01
a.u. $\approx$ 270 meV.  The modified field-free bands of this model
are shown in Fig.~\ref{fig:bandclosing1}~(a), in grey (and also in
Fig.~\ref{fig:bandclosing1}~(b), which is a cut of
Fig.~\ref{fig:bandclosing1}~(a) along direction $y$).  We have then
performed an optimization with the target of closing the gap at point
$K$: for that purpose, we used the functional defined in
Eq.~(\ref{eq:functionalF}), using only the $K$ point to define the set
$\mathcal{K}$, and setting the target pseudoenergies of both bands to
zero. The resulting optimal pseudobands are shown in both
Figs.~(\ref{fig:bandclosing1})~(a) and (b): it can be
seen how the optimization is successful and the gap is effectively
closed.

\begin{figure}
\centerline{\includegraphics[width = 0.9\columnwidth]{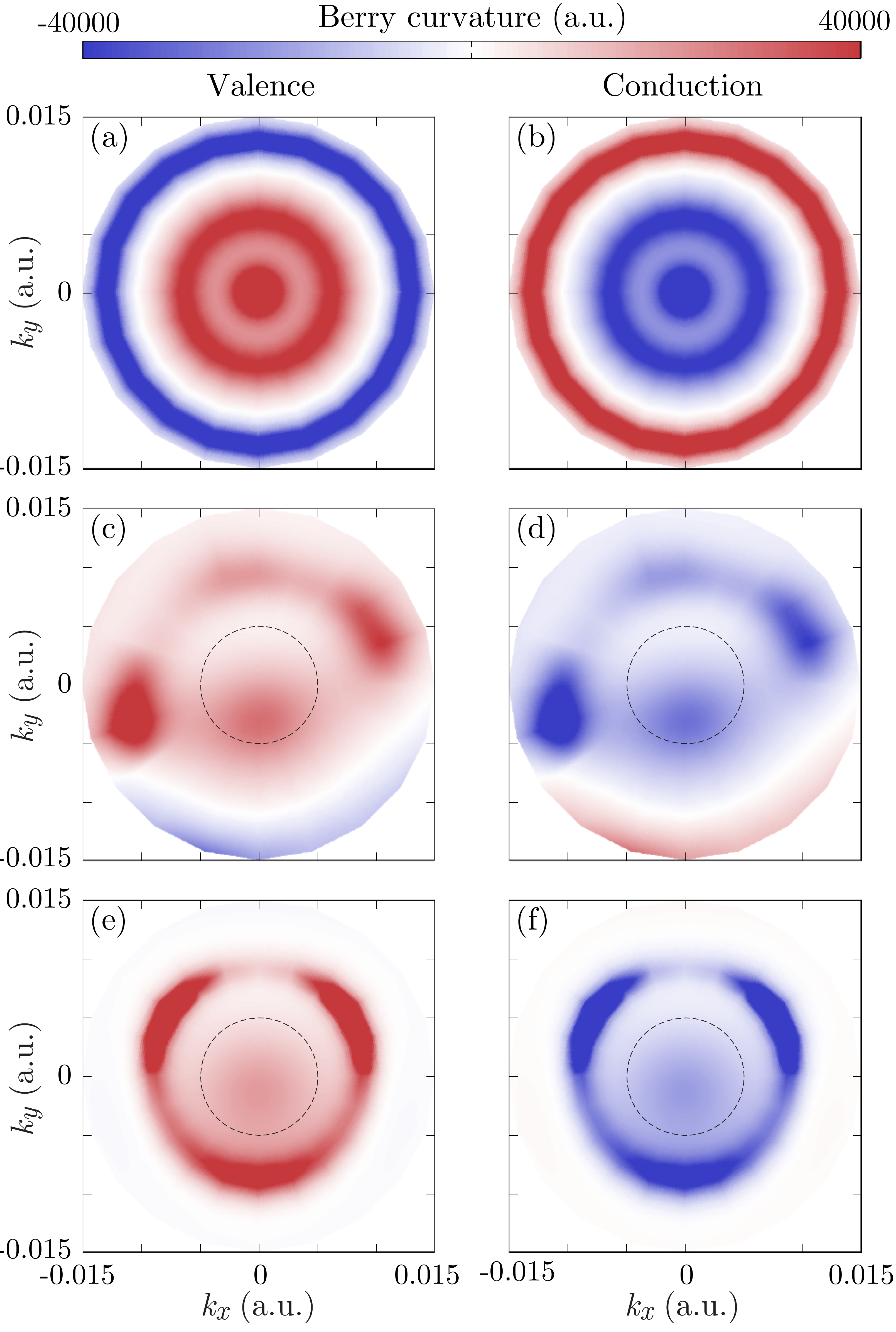}}
\caption{\label{fig:berry}
Berry curvature of the valence (left panels) and conduction (right panels) pseudo-bands. They correspond to: (top panels: a, b) the Floquet pseudo-bands computed with the reference circulary polarized fields; (medium panels: c, d) the flat pseudo-bands; and (bottom panels: e, f) the negative curvature bands shown in Fig.~\ref{fig:effectiveMass}. The top panel (a) replicates the results shown in Ref.~\cite{Sato2019a} [Fig. 3(d)].
Black dashed circles indicate the area for which the target was defined.}
\end{figure}

We now discuss how the flexible band engineering by OCT shown above is
accompanied by changes to the Berry curvature. In the present work, we
chose not to design pulses that target specific Berry curvatures or
topologies, although this is in principle possible. However, for
completeness, we show in Fig.~\ref{fig:berry} the Berry curvature for
the Floquet pseudo-bands obtained with the optimal pulses shown above
(and also with the reference monochromatic circularly polarized
fields), computed as:
\begin{equation}
\Omega_ {\rm B}({\bf k}) = \frac{-i}{T}\int_0^T\!\!{\rm d}t\;
\left[
\nabla_{\bf k} \times \langle u_{\bf k}^{\alpha} (t) \vert \nabla_{\bf k}\vert u_{\bf k}^{\alpha} (t) \rangle
\right]_z\,.
\end{equation}

Here, $\alpha = 0,1$ for the lower (valence) and higher (conduction)
pseudo-energy Floquet pseudo-band, respectively.  We display them
separately on the left [(a), (c), and (e)] and right [(b), (d), and
  (f)] panels of Fig.~\ref{fig:berry}.  It can be seen that, while the
Berry curvature of the Floquet states obtained with the monochromatic
circular pulse varies strongly, the curvatures of the pseudo-bands
obtained with the two optimized cases show more gradual changes. It
can also be seen how valence and conduction curvatures take
approximately inverse values. We note that a region of the BZ outside
the target area is also shown; inside the target area the Berry
curvature is very small in both cases. The variations seem to be the
result of small changes of the optimal results. This indicates the
sensitivity of the Berry curvature to the shape of the bands, and to
the driving protocol. This would hint that an optimization protocol
that would directly target the Berry curvature could be very efficient
in manipulating this property.

\subsection{Design of the switch-on pulse}

\begin{figure}
\includegraphics[width = \columnwidth]{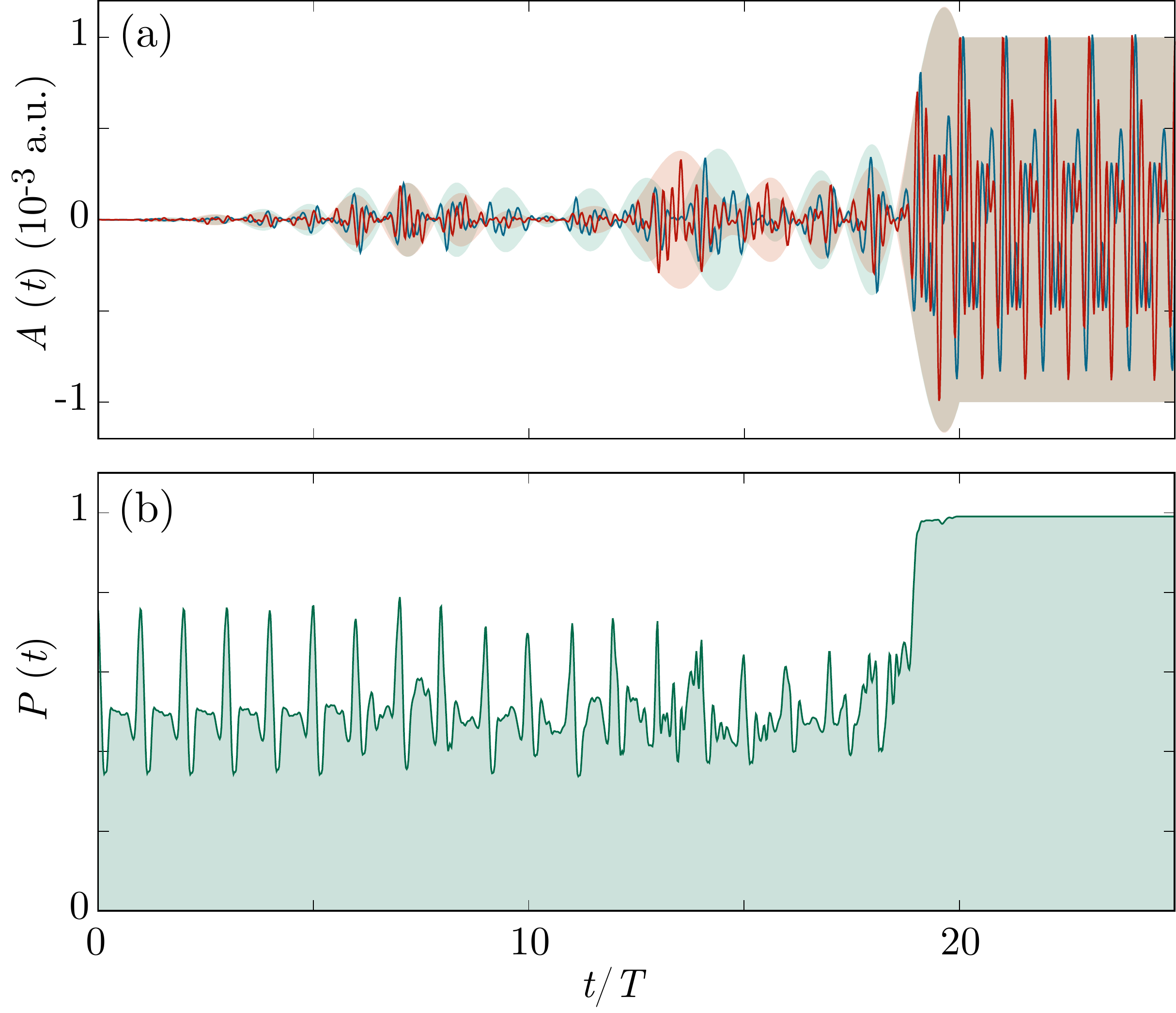}
\caption{\label{fig:switchon}
(a) Optimal switch-on $A^{\rm so}_x$ (blue) and $A^{\rm so}_y$ (red) pulses,
applied during the first 20 first periods of duration $T$, followed by the optimized periodic drivings 
$A_x$ (solid red) and $A_y$ (dashed red). Lines represent the full pulses $A^{\rm so}_i(v^{\rm opt}, t) = S(t) f_i(v^{\rm opt}, t) A_i(u^{\rm opt}, t)$,
whereas the filled curves are the absolute value of the envelopes $S(t) f_i(v^{\rm opt}, t)$ (which
become equal to one after the switch-on phase). (b)Average
population of the lower pseudo-energy Floquet modes: 
$\vert\langle\psi_{{\bf k}0}(t)\vert u^0_\mathbf{k}(t)\rangle\vert^2$. One
can see how, after the switch-on phase, it is approximately constant and equal to one 
since the system has been (almost) transferred
to those states.
}
\end{figure}

For the Floquet engineering of materials to be practically useful,
one also needs to be in full control of the occupation of bands:
the material properties not only depend on the band shapes, but on how they are occupied. Therefore, before the periodic
drivings are started, there should be a switch-on phase in which a pulse leads the electrons in the valence bands to occupy
the Floquet bands of interest. OCT can be used to design this switch-on pulse, and we show in Fig.~\ref{fig:switchon} one example,
demonstrating that it is possible to target the full occupation of a single Floquet state. In particular, this plot corresponds
to the problem of design of flat bands. The target is to transfer the electrons in the equilibrium valence band to the lower
Floquet pseudoband that was found via the previous optimization process (in the target region defined previously,
a disk around the $K$ point of the BZ).

As discussed in the previous section, we have enforced some constraints on the switch-on pulse, since we are aware that the experimental realization of these ideas would also face technical constraints. For this proof-of-principle study, we have envisioned a switch-on phase that consists of the same periodic drives that are later responsible for the tailored bands, but multiplied by a smooth envelope that slowly ramps from zero to one. This envelope admits some low frequencies, and the amplitudes of these are the control parameters in the optimization process. One also needs to specify a duration for the switch-on phase; in our case we have set it to 20 times the fundamental period of the periodic drive $T \approx 21.7$~fs. The top panel of Fig.~\ref{fig:switchon} displays the resulting optimal pulses; the thin lines are the full electric fields [Eq.~(\ref{eq:switchonpulses})], whereas the shaded filled curves are the envelopes. One can see how they smoothly morph into the periodic drives. Of course, the true experimental constraints regarding the maximal frequencies or amplitudes may vary, but the OCT methodology that we have described may be adapted accordingly.

To illustrate the evolution of the Floquet occupations under the
optimized pulses, we show the average of the square modules of the
projection of the time evolving states onto the Floquet states $\vert
u^0_\mathbf{k}(t)\rangle$ (``0'' is the index of the lower
pseudo-band):
\begin{equation}
P(t) = \frac{1}{N_{\rm kps}}\sum_{k\in\mathcal{K}}
\vert\langle\psi_{{\bf k}0}(t)\vert u^0_\mathbf{k}(t)\rangle\vert^2\,.
\end{equation}
By design, the switch-on pulse stops once the occupation has reached unity, after which the periodic Floquet driving is activated and the occupation remains (approximately) constant, as intended.

\section{Conclusions}
\label{sec:conclusions}

We have demonstrated, by performing proof-of-principle OCT calculations, that one can create and control the non-equilibrium steady state Floquet phase with an unprecedented range of versatility, by finding the necessary multi-frequency periodic drivings. The Floquet band structure of materials can thus be shaped at will over the regions of the Brillouin zone deemed to be relevant in each case. This has strong implications on optical and transport properties. We have not used global or local topological properties as optimization targets, but the sensitivity of the Berry curvature shown here, clearly indicates that manipulation of this and other topological properties is possible. Thus, this proposed design of driving pulses is an important step towards designing quantum materials ``on demand''~\cite{Hsieh:2017ix}. Furthermore, we have not only optimized the Floquet band structure shape by designing the periodic drivings, but also demonstrated the possibility of controlling the occupation of Floquet states via a switch-on pulse optimization. Here we have assumed a closed quantum system, and employed a single-electron tight-binding model. However, the case of correlated materials can be treated similarly and is left for a future work. The results shown in the present work are naturally transportable to the case of including electron correlations in the Hamiltonian.

Finally, the discussed scheme also offers a route towards controlling the stability of Floquet states and affecting their long time behaviour. It is expected that in the early times a system will absorb significantly less energy when prepared as a single Floquet eigenstate, because the entire dressed state, electronic structure and laser, is an eigenstate of the time-dependent Schr\"odinger equation. However, decoherence, dissipation effects, and scattering pathways that are not included in the Schr\"odinger equation can affect this state. Therefore, the state will eventually lose its pure Floquet character and gradually heat up to $T\rightarrow\infty$ at infinite times. However, the so called Floquet pre-thermal state may last much longer for a pure Floquet state, thus making it accessible for technological applications. Additionally one could envision OCT schemes that account for the major decoherence channels and thereby design pulses that counteract them, and thus increase the Floquet states lifetimes. The extension of the current work to deal with these issues by considering open quantum systems is work in progress.

\section{Acknowledgements}
We acknowledge support by the Cluster of Excellence ``Advanced Imaging of Matter'' (AIM), Grupos Consolidados (IT1249-19) and Deutsche  Forschungsgemeinschaft (DFG) – SFB-925 – project 170620586. 
The Flatiron Institute is a division of
the Simons Foundation. AC acknowledges support from the AEI grant
FIS2017-82426-P.

\bibliography{bib}
\end{document}